\documentclass[aps,prl,twocolumn,preprintnumbers]{revtex4-2}
\usepackage{amsmath, mathrsfs, amssymb,amsfonts,amsthm,graphicx, epsf, dcolumn, yfonts, subfigure}
\usepackage[hyperfootnotes=true]{hyperref}
\usepackage{color}
\usepackage{slashed}
\usepackage{setspace}
\usepackage{cancel}
\usepackage{wasysym}
\usepackage{bm}
\usepackage{float}
\usepackage{mathtools}
\usepackage{enumerate}
\pdfoutput=1
 \newcommand{\be}{\begin{equation}}
 \newcommand{\ee}{\end{equation}}
 \newcommand{\bea}{\begin{eqnarray}}
 \newcommand{\eea}{\end{eqnarray}}

\newcommand{\beq}{\begin{equation}}
\newcommand{\eeq}{\end{equation}}

\renewcommand*{\thefootnote}{\fnsymbol{footnote}}

\newcommand{\overbar}[1]{\mkern 1.5mu\overline{\mkern-1.5mu#1\mkern-1.5mu}\mkern 1.5mu}

\begin{document}

%\preprint{IFT-UAM/CSIC-23-126}

\title{Gravitational R\'enyi entropy from corner terms}
\author{Jani Kastikainen$^{1}$ and Andrew Svesko$^{2}$}
\affiliation{$^1$Institute for Theoretical Physics and Astrophysics and Würzburg-Dresden Cluster of Excellence
ct.qmat, Julius-Maximilians-Universität Würzburg, Am Hubland, 97074 Würzburg, Germany\\
$^2$Department of Mathematics, King’s College London, Strand, London, WC2R 2LS, UK}

\begin{abstract}\vspace{-2mm}
\noindent We provide a consistent first principles prescription to compute gravitational R\'enyi entropy using Hayward corner terms. For Euclidean solutions to Einstein gravity, we compute R\'enyi entropy of Hartle--Hawking and fixed--area states by cutting open a manifold containing a conical singularity into a wedge with a corner. The entropy functional for fixed--area states is equal to the corner term itself, having a flat-entanglement spectrum, while extremization of the functional follows from the variation
of the corner term under diffeomorphisms. Notably, our method does not require regularization of the conical singularity, and naturally extends to higher-curvature theories of gravity.

\end{abstract}

\renewcommand*{\thefootnote}{\arabic{footnote}}
\setcounter{footnote}{0}

\maketitle

\noindent \textbf{Introduction.} Gravity has an information theoretic character. Evidence for this is captured by 
the Ryu--Takayanagi prescription for computing entanglement entropy of holographic conformal field theories (CFT) \cite{Ryu:2006bv,Ryu:2006ef},
\beq S_{\text{EE}}(A)=\underset{\mathcal{C}\,\sim\,\partial A}{\text{min}}\frac{\text{Area}\,(\mathcal{C})}{4G_{\text{N}}}\;.\label{eq:RTform}\eeq
Namely, entanglement entropy of a holographic CFT state reduced to a (spatial) subregion $A$ of the boundary of anti-de Sitter (AdS) space equals the area of a bulk minimal surface $\mathcal{C}$ anchored to boundary $\partial A$ and homologous to $A$. The relation (\ref{eq:RTform}) generalizes the Bekenstein--Hawking entropy formula~\cite{Bekenstein:1972tm,Bekenstein:1973ur,Hawking:1975vcx,Hawking:1976de}, revealing surfaces other than horizons carry entropy. Further, a possible microscopic interpretation of gravitational entropy is that it measures entangled degrees of freedom of a dual CFT. 
%\JK{I think we should not refer to bulk here since the idea is to first construct the replica CFT manifold and then extend that into the bulk. We can just write: "To wit, consider a holographic CFT on a manifold $B$.}

The  prescription (\ref{eq:RTform}) has a well-known derivation at the level of the gravitational path integral \cite{Lewkowycz:2013nqa}. To wit, consider a CFT living on the boundary $B$ of AdS.
%consider a bulk Riemannian manifold $(\mathcal{M},\overbar{g})$ endowed with Euclidean metric $\overbar{g}$ and asymptotic boundary~$B$.
 %$(B,\gamma)$ with metric $\gamma$. 
 Then invoke the `replica trick': glue together integer $n$-copies of~$B$, producing an $n$-fold cover~$B_{n}$ with partition function $Z[B_n]$. The entanglement entropy of a quantum state reduced to a boundary subregion $A$ is given by the  analytic continuation $n\to1$ of the $n$th R\'enyi entropy,
 %\JK{Would notation $Z_B[\gamma]$ be too messy where $\gamma$ is the ads boundary metric?}
 \beq S_{\text{EE}}(A)=-\partial_{n}(\log Z[B_{n}]-n\log Z[B])|_{n=1}\;.\label{eq:HEEgenCFT}\eeq
 Via AdS/CFT, the boundary partition function $Z[B_{n}]$ may be evaluated in the saddle-point approximation by the on-shell action $I_{\mathcal{M}_{n}}[\overbar{g}_{n}]$ of a regular bulk solution $(\mathcal{M}_n,\overbar{g}_{n})$ to the bulk field equations with boundary $B_n$.
 %Via AdS/CFT, the boundary partition function $Z[B_{n}]$ may be evaluated using the bulk partition function $Z[\mathcal{M}_n]$. Classically, $Z[\mathcal{M}_n]$ is given by the on-shell action $I_{\mathcal{M}_{n}}[\overbar{g}_{n}]$, where $\overbar{g}_{n}$ is a smooth solution to the bulk field equations. 
 %,\JK{ taken to be the boundary of a bulk Riemannian manifold~$(\mathcal{M}_{n},g_{n})$. After (2)?}
  For Hartle--Hawking states (defined below) and assuming $\overbar{g}_{n}$ preserves the $\mathbb{Z}_{n}$ permutation symmetry of $n$-replicas, the entropy (\ref{eq:HEEgenCFT}) can be computed  by
 %of the orbifold $\hat{\mathcal{M}}_{n}\backslash\mathcal{C}$ \JK{Also drop $\hat{\mathcal{M}}_n$ notation? I would call $(\mathcal{M}_n,\overbar{g}_n)$ regular instead of smooth and say $(\mathcal{M}_n/\,\mathbb{Z}_n,\overbar{g}_n)$ has a conical singularity.}
 \beq S_{\text{EE}}(A)=\partial_{n}I_{\mathcal{M}_{n}/\,\mathbb{Z}_{n}}[\overbar{g}_n]\big|_{n=1}\;.\label{eq:entfuncaction}\eeq
 Here the orbifold $(\mathcal{M}_{n}/\,\mathbb{Z}_{n},\overbar{g}_{n})$ is regular everywhere except along a bulk codimension-2 surface $\mathcal{C}$ with a conical defect due to the fixed points of $\mathbb{Z}_{n}$. 
 %For Einstein gravity, the right hand side (\ref{eq:entfuncaction}) returns (\ref{eq:RTform}), where $\mathcal{C}$ is dynamically determined to be a minimal surface.
 For Einstein gravity, (\ref{eq:entfuncaction}) returns (\ref{eq:RTform}), where the area and its minimization are computed in the solution $(\mathcal{M},\overbar{g})$ with boundary $B$ \cite{Lewkowycz:2013nqa}.%\JK{We should say that the area is computed and the minimization is done in the original on-shell metric $\overbar{g}$ dual to $B$.}\JK{Possible reference \cite{Brustein:2021qkj} to talk about removal of $\mathcal{C}$.}

In this letter we present a new method of deriving gravitational entropy using a technique we call the `corner method'. Key to our approach is to recognize the conical singularity arising from the replica trick as a corner: a codimension-2 surface at the intersection of two codimension-1 boundaries. We cut open the conical singularity into a wedge whose boundaries meet at a corner, such that a Hayward corner term is required to have a well-posed variational problem \cite{Hayward:1993my}. This cutting has no effect on the value of the gravitational action, such that the Euclidean action of the wedge entirely encodes the gravitational entropy functional, and is consistent with the extremization prescription. Alternatively, our observation provides a rigorous \textit{definition} of the action of a conical singularity that does not require regularization.

Historically, corner terms have been used to compute entropy of stationary black holes \cite{Banados:1993qp,Hawking:1994ii,Teitelboim:1994az,Teitelboim:1994is}.  i.e., solutions whose Euclideanization have a $U(1)$ Killing symmetry with the bifurcate Killing horizon being a fixed point of the $U(1)$ isometry. Our approach thus extends these computations to backgrounds without a $U(1)$ symmetry, analogous to how \cite{Lewkowycz:2013nqa} generalizes the Gibbons--Hawking derivation of black hole thermodynamics \cite{Gibbons:1976ue}.

There are three notable features of our approach. First, we directly compute entropy functionals and their extremization for Hartle--Hawking and fixed--area states.  For fixed--area states, area minimization follows from varying the Einstein action of the wedge under transverse diffeomorphisms of the corner. Second, unlike derivations \cite{Lewkowycz:2013nqa,Fursaev:2006ih}, we need not regularize any conical singularity. Thirdly, our method extends to higher-curvature theories. In Lovelock gravity, for example, fixed--area states generalize to fixed Jacobson--Myers functional states, having a flat entanglement spectrum \cite{Kastikainen:2023omj}. %Moreover, extremization of entropy functionals is consistent with extremization of the wedge action, resolving tension between which prescription is correct \cite{Chen:2013qma,Bhattacharyya:2014yga,Erdmenger:2014tba}.\JK{This last sentence is a bit subtle, because it is related to the analytically continued metrics. We will not really touch on it in the long paper, but rather in the paper about conical singularities.}

\noindent \textbf{Set-up and gravitational states.} While motivated by AdS/CFT, our approach applies more broadly. Let $(\mathcal{M},g)$ be a $D$-dimensional Riemannian manifold endowed with a Euclidean metric $g$, and $(B,\gamma)$ 
be its $(D-1)$-dimensional boundary 
with topology $ B =S^1\times Y$ and metric $\gamma$. Another codimension-1 manifold $(B_n,\gamma)$ is constructed by cutting and cyclically pasting together positive integer $n$-copies of $(B,\gamma)$ along $Y$. We will look for bulk solutions $(\mathcal{M}_n,\overbar{g})$ with boundary $(B_n,\gamma)$. Further, let $\tau\in (0,2\pi)$ be the Euclidean time coordinate parametrizing the circle $S^1$. Then the cutting-gluing surgery extends this range to $\tau\in(0,2\pi n)$, keeping the metric $\gamma$ fixed in these coordinates. Moreover, $B_{n}$ has a $\mathbb{Z}_{n}$ replica symmetry owed to the cyclical gluing.

We define the (refined) gravitational R\'enyi entropy
\begin{equation}
    \widetilde{S}_n \equiv (n\partial_n - 1)\,I_{\mathcal{M}_{n}}[\overbar{g}]\;.
\label{eq:refinedRenyi}\end{equation}
This entropy is related to the standard gravitational R\'enyi entropy $S_{n}$ via $\widetilde{S}_{n}\equiv n^{2}\partial_{n}[(n-1)\,S_{n}/n]$ \cite{Dong:2016fnf}.
To determine the solution $\overbar{g}$, we must specify the gravitational state. We are interested in two types of states:
\begin{enumerate}[(i)]
    \item \emph{Hartle--Hawking (HH) states}, prepared by a Euclidean gravity path integral over all metrics with fixed asymptotics at infinity.
    \item \emph{Fixed--area states}, prepared by a Euclidean gravity path integral over metrics with a given fixed area $\mathcal{A}$ on a codimension-2 surface $\mathcal{C}$ in the interior and fixed asymptotics at infinity.
\end{enumerate}
More carefully, a fixed--area state is defined as follows \cite{Dong:2018seb}. Gauge-fix a Cauchy slice $\Sigma$ such that it passes through a codimension-2 surface $\mathcal{C}$ and fixes the location of $\mathcal{C}$ on $\Sigma$. This defines a state with $\mathcal{C}$ of fixed area  $\mathcal{A}$. The associated bulk wavefunctional is found by restricting the Hartle-Hawking wavefunctional characterizing the HH state on a $\Sigma$ which gives $\mathcal{C}$ area $\mathcal{A}$. In the path integral this amounts to fixing the induced metric on $\Sigma$ to yield area $\mathcal{A}$ for $\mathcal{C}$ in addition to the same asymptotics.

Thus, gravitational states (i) and (ii) induce different boundary conditions at the codimension-2 surface $\mathcal{C}$ where the Euclidean time circle shrinks to zero size. In particular,  let $\rho > 0$ be a radial coordinate where $\mathcal{C}$ is located at $\rho = 0$. For $\tau \sim \tau + 2\pi n$, the  boundary conditions associated to states (i) and (ii) result in the following metric expansions near $\mathcal{C}$:
%%\begin{equation}
%	ds^{2} = ds^{2}_{0} + \Delta (ds^2)
%	\label{fullmetric}
%\end{equation}
%where $ \Delta (ds^2) $ denotes terms that are subleading with respect to $ ds^{2}_{0} $ near $ \mathcal{C} $. There have been two proposals for the leading term.
\begin{enumerate}[(i)]	
	\item Fixed-periodicity boundary condition
	\begin{equation}
		ds^2  =d\rho^{2} +\frac{\rho^{2}}{n^2}\,d\tau^{2}+ \sigma_{AB}(\hat{x})\,d\hat{x}^Ad\hat{x}^B+\ldots
  %2V_A(\hat{x})\,\rho^2d\tau d\hat{x}^A+...
		\label{maldacena}
	\end{equation}
	This boundary condition is used in the proof of prescription (\ref{eq:RTform}) by Lewkowycz and Maldacena \cite{Lewkowycz:2013nqa}.
 \item Fixed-area boundary condition
	\begin{equation}
		ds^2  =d\rho^{2} +\rho^{2}d\tau^{2} + \sigma_{AB}(\hat{x})\,d\hat{x}^Ad\hat{x}^B+\ldots
  %+2V_A(\hat{x})\,\rho^2d\tau d\hat{x}^A+\ldots,
		\label{fursaev}
	\end{equation}
	with 
 the area $\int_{\mathcal{C}} d^{D-2}x\sqrt{\sigma} \equiv \mathcal{A}$ fixed. 
\end{enumerate}
The ellipsis denote subleading terms in $\rho$ and $\hat{x}^A$ are $D-2$ worldvolume coordinates of $\mathcal{C}$. A metric $g_n$ obeying (\ref{maldacena}) has no conical singularity, while a metric $g_{\mathcal{A}} $ with (\ref{fursaev}) has a conical excess.
 The fixed-area condition was used in Fursaev's attempted proof of (\ref{eq:RTform}) \cite{Fursaev:2006ih}, but was shown to have a flat spectrum \cite{Headrick:2010zt}. Roughly speaking, the relation between Hartle-Hawking and fixed-area states is analogous to applying a Legendre transformation (via a Hayward term) and amounts to switching fixed-periodicity boundary conditions to the fixed-area boundary conditions.  This is akin to the transformation between canonical and microcanonical thermal ensembles \cite{Dong:2018seb}, where a fixed-area state is like a thermal state in a microcanonical ensemble with flat entanglement spectrum.

Given boundary conditions (i) or (ii), the bulk field equations, in principle, can be solved for $\overbar{g}_n$ and $\overbar{g}_{\mathcal{A}}$ order by order in proper distance away from $\mathcal{C}$. In practice, however, extracting the on-shell form of the subleading terms in the HH state (\ref{maldacena}) for $n>1$ is non-trivial and known as the \textit{splitting problem} \cite{Miao:2014nxa,Miao:2015iba,Camps:2016gfs}. We review its resolution in Einstein gravity in \cite{Kastikainen:2023omj}.
%\JK{We should say the splitting problem only arises for HH states when $n > 1$.}
There is no splitting problem for fixed--area states (as there is no $n$ in (\ref{fursaev})).

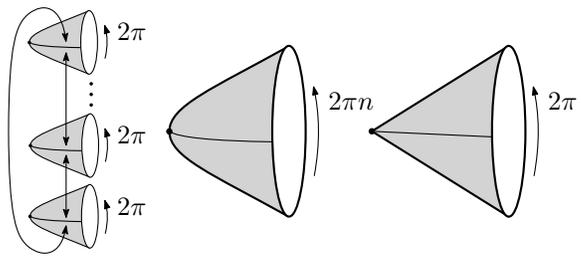
\begin{figure}[t!]
\centering
\begin{center}
\includegraphics{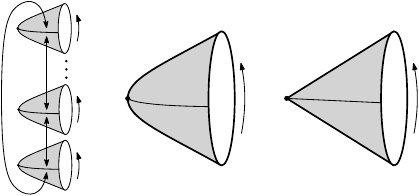}
\put(-160,81){$2\pi$}
\put(-160,42){$2\pi$}
\put(-160,15){$2\pi$}
\put(3,55){$2\pi$}
\put(-80,55){$2\pi n$}
%\put(-201,56){\footnotesize $\mathcal{M}$}
%\put(-221,80){\footnotesize $\partial\mathcal{M}$}
\caption{Left: Replicated $n$-manifold $(\mathcal{M}_n,\overbar{g}_{\mathcal{A}})$ preparing fixed--area states. Center: Manifold $(\mathcal{M}_n,\overbar{g}_{n})$ preparing Hartle--Hawking states. Right: The quotient geometry $(\mathcal{M}_{n}/\,\mathbb{Z}_{n},\overbar{g}_{n})$ for computation of HH entropy.}\vspace{-7mm}
\label{fig:gravstates}
\end{center}
\end{figure}

For states (i), the (proper) circumference of the Euclidean time circle is fixed to be $2\pi$ such that the manifold $(\mathcal{M}_{n},g_{n})$ near $\mathcal{C}$ is regular (without an angular deficit). 
%From a solution $\overbar{g}_n$ the on-shell value of the induced metric $\sigma_{AB}$ of $\mathcal{C}$ is denoted $\overbar{\sigma}_{n}$.
From a solution $\overbar{g}_n$ the on-shell value $\overbar{\sigma}_{n}$ of the induced metric $\sigma$ of $\mathcal{C}$ is determined. Meanwhile, for states (ii), the area of $\mathcal{C}$ is fixed and the bulk near $\mathcal{C}$ is locally a replicated manifold $(\mathcal{M}_{n},g_{\mathcal{A}})$  with conical singularity of angular \emph{excess} $\Delta \tau = 2\pi\,(n-1)$ at $\mathcal{C}$. See Fig. \ref{fig:gravstates}.
%, since $\tau \sim \tau + 2\pi n$.
%The solution $\overbar{g}_{\mathcal{A}}$ depends on $\sigma$ through the area $\mathcal{A}$ of $\mathcal{C}$.  Observe fixed--area metrics are locally $n = 1$ HH metrics.

%A way of distinguishing boundary conditions (i) and (ii) is that 
The entropy of each state is computed differently. First consider Hartle--Hawking states obeying boundary condition \eqref{maldacena}. %and on-shell induced metric by $\overbar{\sigma}_n$.
Now assume the on-shell metric $\overbar{g}_n$ preserves replica symmetry \cite{Lewkowycz:2013nqa}, via $\tau \rightarrow \tau + 2\pi k$ for integer $k\geq 1$. To perform the analytic continuation of $n$, it is useful to work with the orbifold $\mathcal{M}_{n}/\,\mathbb{Z}_{n}$, where now $\tau$ has identification $\tau \sim \tau + 2\pi$. Equally, $\mathcal{M}_{n}/\,\mathbb{Z}_{n} = \mathcal{M}\,\backslash\, \mathcal{C} $ with $\tau \sim \tau + 2\pi$ and $\mathcal{C}$ removed.\footnote{A remark on notation. It is common practice to write $I_{\mathcal{M}_n}[\overbar{g}_n]$ on the right-hand side of Eq. (\ref{eq:refinedRenyi}). We prefer $I_{\mathcal{M}_{n}\backslash\,\mathcal{C}}[\overbar{g}_{n}]$  because we treat $(\mathcal{M}_{n},\overbar{g}_{n})$ topologically as a punctured disc (integrals range over $\rho\in(0,\infty)$, the $\rho=0$ surface is removed). This is consistent with \cite{Dong:2013qoa}.} Thence, $I_{\mathcal{M}_n}[\overbar{g}_n] = n\,I_{\mathcal{M}\,\backslash\, \mathcal{C}}[\overbar{g}_n]$, and the entropy \eqref{eq:refinedRenyi} becomes
\begin{equation}
	\widetilde{S}_n = n^{2}\partial_{n}\,I_{\mathcal{M}\,\backslash\, \mathcal{C}}[\overbar{g}_n],\quad (\text{HH state}).
	\label{refinedrenyiHH}
\end{equation}
As the integration region on $\mathcal{M}\,\backslash\, \mathcal{C}$ is independent of $n$, the $n$-derivative acts only on components of $\overbar{g}_n$, i.e., an on-shell variation of the action. This requires an analytic continuation of the on-shell action to non-integer $n$.

Alternatively, the solution $\overbar{g}_\mathcal{A}=\overbar{g}_{1}$ obeying the fixed-area condition \eqref{fursaev}  is independent of $n$ but depends on the area $\mathcal{A}$ of $\mathcal{C}$. Thus, the Rényi entropy (\ref{eq:refinedRenyi}) is
\begin{equation}
	\widetilde{S}_n = (n\partial_n-1)\,I_{\mathcal{M}_n}[\overbar{g}_\mathcal{A}],\quad (\text{fixed-area state}),
	\label{refinedrenyifixedarea}
\end{equation}
with $\partial_{n}$ only acting on the integration region $\mathcal{M}_{n}$. %since $\overbar{g}_{\mathcal{A}} = \overbar{g}_1$ is independent of $n$. 
%\JK{Could be mentioned before.} 
%This differs from the quotient manifold $(\mathcal{W}_1,g_n)$ of the previous section which has a conical deficit.

A distinction between the R\'enyi entropies of the Hartle--Hawking and fixed--area states is that the former is $n$-dependent while the latter is $n$-independent. Hence, the entanglement spectrum of a fixed--area state is `flat' \cite{Akers:2018fow,Dong:2018seb}:
%the eigenvalues of the modular Hamiltonian of its associated density operator are sharply peaked \cite{Akers:2018fow,Dong:2018seb},\JK{This sharply peaked wording refers to the expansion of the modular Hamiltonian in Newton's constant where the leading term is the area operator. I think mentioning it is confusing.} 
 the reduced density matrix describing the state is
(approximately) proportional to the identity operator. 
%Intuitively, we can think of the Hartle--Hawking state as a thermal state in the canonical ensemble, while the fixed--area state restricts the thermal to a small energy window, and is thus a state in the microcanonical ensemble.

%The gravitational entropy of fixed--area states has been explicitly demonstrated for Einstein gravity \cite{Dong:2018seb}, and AdS Jackiw-Teitelboim (JT) gravity \cite{Arias:2021ilh}. Using corner terms, we will upgrade this derivation in the case of Lovelock gravity. Notably, in a first attempt at proving the Ryu-Takayanagi formula, Fursaev computed an $n$-independent R\'enyi entropy \cite{Fursaev:2006ih}, though his derivation was not accepted as a complete proof of the RT formula as it did not match with the CFT calculation of R\'enyi entropy in $\text{AdS}_3\slash \text{CFT}_2$ \cite{Headrick:2010zt}.

\begin{figure}[t!]
\begin{center}
\hspace{-10mm}\includegraphics{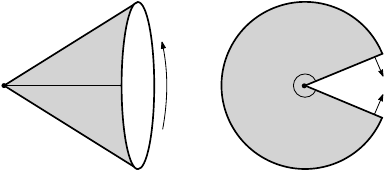}
\put(-197,39){ $\mathcal{C}$}
\put(-150,31){{$\mathcal{B}$}}
\put(-176,4){ $\mathcal{M}\,\backslash\,\mathcal{C}$}
\put(-105,39){ $2\pi$}
\put(-50,10){$\mathcal{W}$}
\put(-28,55){{$\mathcal{B}_{1}$}}
\put(-28,24){{$\mathcal{B}_{2}$}}
\put(2,32){$n_{2}$}
\put(2,45){$n_{1}$}
%\put(-4,31){\footnotesize {$n_{2}$}}
%\put(-4,41){\footnotesize {$n_{1}$}}
%\put(-45,38){\footnotesize {$r_{1}$}}
%\put(-45,46){\footnotesize {$r_{2}$}}
\caption{Cutting the manifold $(\mathcal{M}_n/\,\mathbb{Z}_n,g_n)$ into a wedge.}\vspace{-9mm}
\label{fig:wedgecone}
\end{center}
\end{figure}
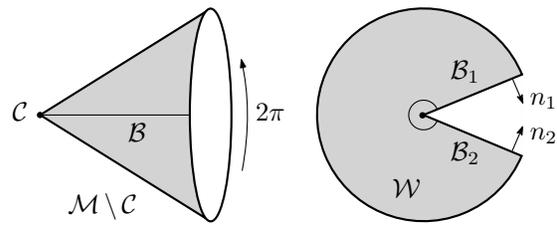

\noindent \textbf{Entropy of Hartle--Hawking states.} 
%Let us now compute the R\'enyi entropy for Hartle--Hawking states using the corner method, assuming bulk Einstein gravity.
 Let $g_n$ be an off-shell metric obeying \eqref{maldacena} on $\mathcal{M}\,\backslash \,\mathcal{C}$ with $\tau \sim \tau + 2\pi$. %Off-shell the induced metric $\sigma$ of $\mathcal{C}$ is free, but on-shell it is fixed to the value $\overbar{\sigma}_n$. 
Notice that cutting $\mathcal{M}\,\backslash \,\mathcal{C}$ open along a codimension-1 surface $\mathcal{B}$, such that $\partial \mathcal{B} = \mathcal{C}$, produces a wedge shaped space $\mathcal{W}$ with two boundaries $\mathcal{B}_{\alpha}$ (with $\alpha = 1,2$) meeting at a corner $\mathcal{C} = \mathcal{B}_{1}\cap \mathcal{B}_{2}$. See Fig. \ref{fig:wedgecone}. Boundaries $\mathcal{B}_1$ and $\mathcal{B}_2$ are located at $\tau = 0$ and $\tau = 2\pi$, respectively. We emphasize, as a topological space, we treat $\mathcal{W}$ as an open set such that $\mathcal{B}_{\alpha}$ and $\mathcal{C}$ are not included in $\mathcal{W}$.

Notably, cutting has no effect on the value of the action as it only removes a sliver of measure zero from the integration region on $\mathcal{M}$.
Thus,
\begin{equation}
I_{\mathcal{M}\,\backslash\,\mathcal{C}}[g_n] = I_{\mathcal{W}}[g_n].
\label{cutHH}
\end{equation}
%which states that the gravitational action of the manifold $(\mathcal{M}\,\backslash \,\mathcal{C},g_n)$ is equal to the action of a manifold $(\mathcal{W},g_n)$ with a corner whose opening angle is determined by $ n $. 
The action of the wedge $\mathcal{W}$ is (in units $16\pi G_{\text{N}}=1$)
\begin{equation}
\hspace{-1mm}I_{\mathcal{W}}[g_n] =-\int_{\mathcal{W}}\hspace{-3mm} d^{D}x\sqrt{g_{n}}\,R - \sum_{\alpha=1}^{2}\int_{\mathcal{B}_{\alpha}}\hspace{-3mm}d^{D-1}x\sqrt{h_{\alpha }}\,2K_\alpha.
\label{fullEinsteinaction}
\end{equation}
%\begin{equation}
%I_{\mathcal{W}_n}[g] = -\int_{\mathcal{W}_n}d^{D}x\sqrt{g}\,R - 2\int_{\mathcal{B}_{1}}d^{D-1}x\sqrt{h_{1}}\,K_1- 2\int_{\mathcal{B}_{2}}d^{D-1}x\sqrt{h_{2}}\,K_2
%\end{equation}
  Here $h_{\alpha ij}$ is the induced metric on $\mathcal{B}_{\alpha}$, $K_{\alpha ij} = h_{\alpha i}^k h_{\alpha j}^{l} \nabla_k n_{\alpha l}$ is its extrinsic curvature with outward-pointing unit normal $n_{\alpha}^a$ to $\mathcal{B}_{\alpha}$. 
 Despite $\mathcal{W}$ not including boundaries $\mathcal{B}_{\alpha}$, 
  we have included a Gibbons--Hawking--York (GHY) term on each boundary.
This is allowed to because the induced metrics and extrinsic curvatures of $\mathcal{B}_{\alpha}$ obey (the relative minus sign is due to oppositely pointing normals $n_{1,2}$),
\begin{equation}
    h_{1ab}\lvert_{\mathcal{B}_1}\, = h_{2ab}\lvert_{\mathcal{B}_2},\quad K_{1ab}\lvert_{\mathcal{B}_1}\, = -K_{2ab}\lvert_{\mathcal{B}_2},
    \label{periodicBCnonpert}
\end{equation}
for integer $n\geq 1$ due to replica symmetry $\tau \sim \tau + 2\pi k$ of $g_n$. Thus, the boundary terms cancel in \eqref{fullEinsteinaction} provided $n$ is an integer. 
%Further, boundary conditions (\ref{periodicBCnonpert}) ensure  the metric and its first derivative are periodic and continuous across the cut.
For non-integer $n$, however, replica symmetry is broken, resulting in a discontinuity in the derivative of the metric when identifying the $\tau = 0,2\pi$ surfaces. Thus, we work at integer $n$ and only analytically continue values of on-shell actions at the end.

%i.e., $n_{\alpha}\cdot n_{\alpha}=1$.

Varying the action \eqref{fullEinsteinaction} with respect to the metric without imposing any boundary conditions yields
\begin{equation}
\begin{split}
&\hspace{-1mm}\delta_{g_{n}} I_{\mathcal{W}}= -\int_{\mathcal{W}}\hspace{-2mm}d^{D}x\sqrt{g_{n}}\,G_{nab}\,\delta g_{n}^{ab}+2\int_{\mathcal{C}}\hspace{-1mm}d^{D-2}x\sqrt{\sigma}\,\delta_{g_{n}} \Theta_n\\
&-\sum_{\alpha=1}^{2}\int_{\mathcal{B}_{\alpha}}\hspace{-2mm}d^{D-1}x\sqrt{h_{\alpha }}\,\widetilde{T}_{\alpha ab}\,\delta h^{ab}_{\alpha},
\label{fulleinsteinvariation}
\end{split}
\end{equation}
where $G_{nab}$ is the Einstein tensor, $\widetilde{T}_{\alpha ab} = K_{\alpha ab} - K_{\alpha } h_{\alpha ab}$ is the boundary stress-tensor,
and the corner angle is given by $\cos{\Theta_n} = g_{n ab}\,n_1^an_2^b$. 
Since the embedding of the first boundary is $\tau = 0$ and the second is $\tau = 2\pi$, explicitly we find $\Theta_n = \pi\,(1-2n^{-1})$.%\JK{Should we add references for \eqref{fulleinsteinvariation} including our long paper?}

%\begin{equation}
%\Theta_n = \pi\,(1-2n^{-1}).
%\label{Thetan}
%\end{equation}

Via the identity \eqref{cutHH}, we can obtain expressions for the action on the manifold $(\mathcal{M}\,\backslash\,\mathcal{C},g_n)$ and its variation. 
%With periodic boundary conditions (\ref{periodicBCnonpert}),
Imposing periodic boundary conditions
\beq \delta h_{1}^{ab}|_{\mathcal{B}_{1}}=\delta h_{2}^{ab}|_{\mathcal{B}_{2}},\label{eq:PBCdeltah}\eeq
such that the metric variation is continuous across the cut, the second line of variation (\ref{fulleinsteinvariation}) is cancelled when $n\geq1$ is an integer.
%We impose periodic boundary conditions (\ref{}) on $\mathcal{B}_{1,2}$ of the wedge
%\begin{equation}
%h_{1ab}|_{\mathcal{B}_{1}}= h_{2ab}|_{\mathcal{B}_{2}},\quad K_{1ab}|_{\mathcal{B}_{1}} = -K_{2ab}|_{\mathcal{B}_{2}},
%\label{periodicBCnonpert}
%\end{equation}
%where the relative minus sign is due to oppositely pointing normals $n_{1,2}$. Periodic boundary conditions (\ref{periodicBCnonpert}) ensure  the metric and its first derivative are periodic and continuous across the cut. 
%Further, the codimension-1 boundary terms in 
%\eqref{fullEinsteinaction} and \eqref{fulleinsteinvariation} cancel.
%leaving
%\begin{equation}
%\delta_gI_{\mathcal{M}}[g_n] =-\int_{\mathcal{M}}\hspace{-2mm}d^{D}x\sqrt{g_n}\,G_{n ab}\,\delta g^{ab}  +2\int_{\mathcal{C}}\hspace{-1mm}d^{D-2}x\sqrt{\sigma}\,\delta_g \Theta_n.
%\label{Mvariation}
%\end{equation}
Since we want to extremize the action over metrics satisfying fixed-periodicity boundary condition (i) \eqref{maldacena} at $\mathcal{C}$, the metric variation is such that $\delta n = 0$. Thus, $\delta_{g_{n}} I_{\mathcal{M}\,\backslash\,\mathcal{C}}[g_{n}]=0$ imposes Einstein's equations $ G_{n ab} = 0$ on the metric $g_n$ everywhere outside $\mathcal{C}$. 

Without going into detail (see \cite{Camps:2016gfs,Kastikainen:2023omj}), imposing Einstein's equations provides a condition constraining $\mathcal{C}$. Namely, expanding the Ricci tensor near $\mathcal{C}$ yields
%\beq \overbar{\sigma}_{1}^{AB}\overbar{K}_{(p)AB}=0\;,\label{eq:condC}\eeq
\begin{equation}
\overbar{\sigma}_{1}^{ab}\,\overbar{L}_{\alpha ab} =\overbar{\sigma}_{1}^{ab}\,\overbar{Q}_{\alpha ab} = 0,
\label{eq:condC}\end{equation}
where $\overbar{\sigma}_{1ab}$ is the induced metric and $\overbar{L}_{\alpha ab},\overbar{Q}_{\alpha ab}$ are extrinsic curvatures
\begin{equation}
	L_{ab} = \sigma_{a}^{c}\,\sigma_{b}^{d}\,\nabla_{c}n_{\alpha d}, \quad Q_{ab} = \sigma_{a}^{c}\,\sigma_{b}^{d}\,\nabla_{c}r_{\alpha d}.
 \label{cornerextrinsiccurvatures}
\end{equation}
of $\mathcal{C}$ in the solution $\overbar{g}_{1}$. Here $r_{\alpha}$ is a vector tangent to $\mathcal{B}_{\alpha}$ obeying $r_{\alpha}\cdot r_{\alpha}=1$ and $r_{\alpha}\cdot n_{\alpha}=0$. Thus, $\mathcal{C}$ is constrained to be a minimal area surface in $\overbar{g}_1$, and the variational principle for the HH metric  fixes the embedding of $\mathcal{C}$.
%which was assumed to be fixed from the beginning (fixed at $\rho = 0$).

Let us now determine the entropy functional. 
%The two variational principles above determine the on-shell metric $\overbar{g}_n$ and the on-shell embedding $\overbar{E}_n$ of $\mathcal{C}$ in $\mathcal{M}$. Together they determine the on-shell induced metric $\overbar{\sigma}_n$ of $\mathcal{C}$. 
Working on-shell and using periodic boundary conditions (\ref{eq:PBCdeltah}),
\begin{equation}
\delta_g I_{\mathcal{M}\,\backslash\,\mathcal{C}}[\overbar{g}_n] = 2\int_{\mathcal{C}}d^{D-2} x\sqrt{\overbar{\sigma}_n}\, \delta_g \overbar{\Theta}_{n},
\label{generalHHvariation}
\end{equation}
where $n\geq1$ is an integer to ensure cancellation of boundary terms via \eqref{periodicBCnonpert}. 
To compute the entropy \eqref{refinedrenyiHH}, however, we consider metric variations corresponding to $\delta n$, requiring analytic continuation of $n$ to real values. Extending to non-integer $n$, then using \eqref{generalHHvariation}, the entropy \eqref{refinedrenyiHH} is
\begin{equation}
	\hspace{-1mm}\widetilde{S}_n  = 2n^{2}\int_{\mathcal{C}}\hspace{-1mm}d^{D-2}x\sqrt{\overbar{\sigma}_n}\,\partial_{n}\overbar{\Theta}_{n}=4\pi\int_{\mathcal{C}}d^{D-2}x\sqrt{\overbar{\sigma}_n}\;,%-n^{2}\frac{d}{dn}\,I_{\mathcal{C}}[\overbar{g}_n]
	\label{entropyfromcorner}
\end{equation}
the area functional of the corner $\mathcal{C}$ in $\overbar{g}_n$. Further, the limit $n\to1$ recovers prescription (\ref{eq:RTform}) with minimization determined by Einstein's equations, \eqref{eq:condC}, as in \cite{Lewkowycz:2013nqa}.
%Restoring the factor of $1\slash(16\pi G_{\text{N}})$ to the Einstein--Hilbert action, we obtain the usual $1\slash (4 G_{\text{N}})$ factor of the Bekenstein--Hawking entropy. 

\noindent \textbf{Entropy of fixed--area states.} 
%Let us now move to fixed-area states. 
%\cite{dong_flat_2019}. 
Consider an off-shell metric $g_{\mathcal{A}}$ obeying boundary condition (ii) \eqref{fursaev}.
%with a $\sigma$ whose area is fixed to a constant.
%\begin{equation}
%    \int_{\mathcal{C}}d^{D-2}x\sqrt{\sigma}\equiv \mathcal{A}\ .
%\end{equation}%\,\backslash \,\mathcal{C}
We cut the replicated manifold $(\mathcal{M}_n,g_{\mathcal{A}})$ open, producing a manifold $(\mathcal{W}_n,g_{\mathcal{A}})$ with boundaries $\mathcal{B}_{\alpha}$ (now at $\tau = 0$ and $\tau = 2\pi n$)  meeting at a corner $\mathcal{C}$. The action satisfies
\begin{equation}
I_{\mathcal{M}_n}[g_{\mathcal{A}}] = I_{\mathcal{W}_n}[g_{\mathcal{A}}]+I_{\mathcal{C}}[g_{\mathcal{A}}],%\,\backslash \,\mathcal{C}
\label{cutreplica}
\end{equation}
where $I_{\mathcal{W}_{n}}$ is given by \eqref{fullEinsteinaction}.
%We also include a corner term to have a well-posed variational problem (for Dirichlet boundary conditions) because $(\mathcal{M}_{n},g_{\mathcal{A}})$ has a corner in its interior.
We include a corner term to have a well-posed variational problem for fixed-area boundary conditions as $(\mathcal{W}_{n},g_{\mathcal{A}})$ has a corner in its interior (see below). It can be understood as the energy density supporting the conical excess present on $(\mathcal{M}_n,g_{\mathcal{A}})$.
%(unlike HH states). 

Unlike HH states, Einstein's equations for $g_{\mathcal{A}}$ do not constrain the embedding of $\mathcal{C}$ for fixed-area states (see below).
%\JK{Explain why. The reason is that the term imposing it is proportional to $n-1$ which vanishes at $n = 1$. This could be done by adding the behaviour of the Ricci tensor before (13).}
%\JK{Actually this is explained below so maybe we can say "see below" or explain it here.} 
%equire us to consider a variational problem for the embedding of $\mathcal{C}$ separately.
We thus derive fixed-area state entropy in three steps: (1) variational principle for the metric, (2) variational principle for the embedding of $\mathcal{C}$, and (3) show the entropy functional is the on-shell action of such solutions.

\noindent \emph{Variational principle for fixed--area metrics.} 
The variational principle under area-preserving metric variations on the manifold $(\mathcal{M}_n,g_{\mathcal{A}})$ is not well defined because of the angular excess at $\mathcal{C}$. Indeed, after the cutting procedure \eqref{cutreplica}, the metric variation of the  action includes a term localized at $ \mathcal{C} $ which must be cancelled by the variation of $I_{\mathcal{C}}[g_{\mathcal{A}}]$ to make the variational problem well defined. 
%\JK{Remove: We thus supplement the action with a Hayward term \cite{Hayward:1993my} that provides the necessary energy density to support the conical excess present on the manifold $(\mathcal{M}_n\,\backslash\,\mathcal{C},g_{\mathcal{A}})$}.
This is achieved by the Hayward corner term \cite{Hayward:1993my} %The total action supplemented with the Hayward term is 
%\JK{We could write just the corner term and refer to (16).}
%\begin{equation}
%I'_{\mathcal{W}_n}[g_{\mathcal{A}}] =I_{\mathcal{W}_{n}} -2\int_{\mathcal{C}}d^{D-2}x\sqrt{\sigma}\,(\Theta_{1\slash n} + \pi).
%\label{fullEinsteinactioncorner}
%\end{equation}
\begin{equation}
I_{\mathcal{C}}[g_{\mathcal{A}}] = -2\int_{\mathcal{C}}d^{D-2}x\sqrt{\sigma}\,(\Theta_{1\slash n} + \pi).
\label{fullEinsteinactioncorner}
\end{equation}
 Despite $g_{\mathcal{A}}$ being $n$-independent, the corner angle $\cos{\Theta_{1\slash n}} = g_{\mathcal{A}ab}\,n_{1}^{a} n_{2}^{b}$
is since the embedding of $\mathcal{B}_2$ depends on $n$; the two boundaries are at $\tau = 0$ and $\tau = 2\pi n$, giving
$\Theta_{1\slash n} = \pi\,(1-2n)$. We have also included a corner ``counterterm'' proportional to $ 2\pi \mathcal{A}$ in (\ref{fullEinsteinactioncorner}), with a coefficient uniquely fixed such that the  total corner term vanishes at $n=1$ (when there is no corner). 

%We can see that at $n=1$ the angle does not vanish $\Theta_1 = -\pi$ which is the reason why we have included the area counterterm to the action \eqref{fullEinsteinactioncorner} such that the combination $\Theta_{1\slash n}+\pi = 2\pi\,(1-n)$ vanishes at $n = 1$. The condition of vanishing of the corner term at $n = 1$ fixes the coefficient of the area counterterm uniquely to $2\pi$.

Equation \eqref{cutreplica} allows us to obtain the action of the replicated manifold $(\mathcal{M}_n,g_{\mathcal{A}})$ from the wedge action \eqref{fullEinsteinactioncorner}. Via periodic boundary conditions, the codimension-1 boundary terms in 
\eqref{fullEinsteinaction} cancel for integer $n$, giving
\begin{equation}%\hspace{-4mm}
I_{\mathcal{M}_n}[g_{\mathcal{A}}] =-n\hspace{-1mm}\int_{\mathcal{M}\,\backslash\,\mathcal{C}}\hspace{-5mm}d^{D}x\sqrt{g_{\mathcal{A}}}\,R -4\pi\,(1-n)\hspace{-1mm}\int_{\mathcal{C}}\hspace{-1mm}d^{D-2}x\sqrt{\sigma}\ .\hspace{0.1cm}
\label{replicaactioncorner}
\end{equation}
where we have used replica symmetry to pull out the factor of $n$ in the bulk integral.
We have thus recovered the distributional contribution of a squashed conical excess to the Ricci scalar originally derived in \cite{Fursaev:2013fta}. There a regularization scheme was employed, including regularization dependent terms that enter at higher orders in $ n-1 \ll 1 $ (except in two dimensions or when the extrinsic curvatures vanish). Notably, our corner method does not produce such terms.
%\JK{I think the issue is that their regulator $p$ is tied to $n$. One could treat $p$ as a separate parameter and take the limit $b\rightarrow 0$ and $p\rightarrow 1$ separately.}

We now extremize the action \eqref{replicaactioncorner} over metrics with fixed area at $\mathcal{C}$. 
%This is achieved by considering variations for which the change in the induced metric of $\mathcal{C}$ is traceless $\sigma_{ab}\,\delta \sigma^{ab} = 0$ as it ensures that $\delta \sqrt{\sigma} = 0$ (fixed area). 
%The variation of \eqref{replicaactioncorner} follows from the variation of the wedge action \eqref{fullEinsteinactioncorner}
The variation of \eqref{replicaactioncorner} follows from \eqref{cutreplica} and the variation of the wedge action \eqref{fullEinsteinaction} and the corner term \eqref{fullEinsteinactioncorner}.
%which without imposing any boundary conditions is given by (see Appendix \ref{app:actionvariation})
%\begin{equation}
%\begin{split}
%\hspace{-4mm}\delta_g I_{\mathcal{W}_n}& = -\int_{\mathcal{W}_n}\hspace{-4mm}d^{D}x\sqrt{g}\,G_{ab}\,\delta g^{ab}-\int_{\mathcal{C}}\hspace{-1mm}d^{D-2}x\sqrt{\sigma}\,\widehat{T}_{ab}\,\delta \sigma^{ab}\\
%&-\sum_{\alpha=1}^{2}\int_{\mathcal{B}_{\alpha}}\hspace{-3mm}d^{D-1}x\sqrt{h_{\alpha }}\,\widetilde{T}_{\alpha ab}\,\delta h^{ab}_{\alpha},
%\label{fulleinsteinvariation}
%\end{split}
%\end{equation}
%where the corner stress tensor is $\widehat{T}_{ab} = -(\Theta_{1\slash n}+\pi)\,\sigma_{ab}$. The boundary terms in \eqref{fulleinsteinvariation} again cancel, 
Under periodic boundary conditions we find
\begin{equation}
\begin{split}%\hspace{-5mm}
\delta_g I_{\mathcal{M}_n}=-\int_{\mathcal{M}_n}\hspace{-4.5mm}d^{D}x\sqrt{g_{\mathcal{A}}}\,G_{ab}\,\delta g^{ab}  -\int_{\mathcal{C}}\hspace{-1.5mm}d^{D-2}x\sqrt{\sigma}\,\widehat{T}_{ab}\,\delta \sigma^{ab}
\label{replicaactioncornervariation}
\end{split}
\end{equation}
with corner stress tensor $\widehat{T}_{ab} = -(\Theta_{1\slash n}+\pi)\,\sigma_{ab}$. The second term vanishes for area-preserving (traceless) variations $\delta \sigma^{ab}$ so that the variational principle imposes Einstein's equations $ G_{ab} = 0 $ on $g_{\mathcal{A}}$ everywhere on the replica manifold. So, the Hayward term, which was originally included to make the Dirichlet problem well defined, also makes the fixed-area variational problem well defined.

\noindent \emph{Extremization from variational problem of embedding.} Einstein's equations for $g_{\mathcal{A}}$ do not give constraints on the embedding of $\mathcal{C}$: since $n=1$ in (\ref{fursaev}), the Ricci tensor responsible for constraints (\ref{eq:condC}) is zero. Thus, the variational problem for the embedding in the on-shell metric $\overbar{g}_{\mathcal{A}}$ is to be considered separately.

Recall $\mathcal{M}_n$ consists of $n$-copies of $\mathcal{M}$ cyclically glued together around $\mathcal{C}\subset \mathcal{M}$. Denote the embedding of $ \mathcal{C} $ in $\mathcal{M}$ as $x^a = E^a(\hat{x})$  and define tangent vectors $e^a_A = \partial_A E^a$ that can be used to pull-back tensors to $\mathcal{C}$. Rather than directly varying the embedding, we keep the embedding fixed $\delta e^a_A  = 0 $ and instead vary the background metric in the wedge $\mathcal{W}$ by an infinitesimal diffeomorphism
\begin{equation}
 \delta_\xi g^{ab} = \nabla^{a}\xi^{b}+\nabla^{b}\xi^{a},\quad \xi^{a} =  \xi^{n}n^{a} + \xi^{r}r^{a},%\xi^{n}n^{a} + \zeta^a = \xi^{n}n^{a} + \xi^{r}r^{a} + p^a,
 \label{transversediffeo}
\end{equation}
where we take the vector field $\xi$ to be normal to $ \mathcal{C} $.
%and $r_{\alpha}$ are vectors tangent to $\mathcal{B}_{\alpha}$.
%obeying $r_{\alpha}\cdot r_{\alpha}=1$ and $r_{\alpha}\cdot n_{\alpha} = 0$. 
We assume $\xi$ and its derivative are continuous across the cut so that the diffeomorphism descends to a transverse variation of the embedding of $\mathcal{C}$ in $\mathcal{M}_n$.

The Einstein action restricted to the wedge domain $\mathcal{W}$ satisfies $I_{\mathcal{W}}[\varphi^*g] = I_{\varphi(\mathcal{W})}[g]$ where $\varphi$ is a diffeomorphism and $\varphi^*g$ is the pull-back of the metric. For domain preserving diffeomorphisms, $\varphi(\mathcal{W}) = \mathcal{W}$, this is the familiar statement of diffeomorphism invariance. At the infinitesimal level, $\varphi^a(x) = x^a - \xi^a(x)$, it relates diffeomorphisms of the metric to variations of the embedding of the corner, $\delta_g I_{\mathcal{W}}[g] = \delta_{\mathcal{W}} I_{\mathcal{W}}[g]$, where on the left-hand side $\delta g^{ab} = \delta_\xi g^{ab} $, while on the right-hand side $\delta_{\mathcal{W}}E^{a}=-\xi^{a}\vert_{\mathcal{C}}$
%$\delta E^a(\hat{x}) = \xi^a(E(\hat{x})) $
(since the embedding of $\varphi(\mathcal{C})$ is $\varphi \circ E$).

Under \eqref{transversediffeo}, the induced metric of $\mathcal{C}$ changes as \cite{Bhattacharyya:2014yga}
\begin{align}
\sigma_{c}^{a}\sigma_{d}^{b}\,\delta_{\xi}\sigma^{cd} &= 2\,\xi^n L^{ab}+ 2\,\xi^r Q^{ab}
\label{inducedmetricvariation}
\end{align}
for extrinsic curvatures of $\mathcal{C}$ in $\overbar{g}_{\mathcal{A}}$ (\ref{cornerextrinsiccurvatures}).
%\begin{equation}
%	L_{\alpha ab} = \sigma_{a}^{c}\,\sigma_{b}^{d}\,\nabla_{c}n_{\alpha d}, \quad Q_{\alpha ab} = \sigma_{a}^{c}\,\sigma_{b}^{d}\,\nabla_{c}r_{\alpha d}.
% \label{cornerextrinsiccurvatures}
%\end{equation}
Applying the diffeomorphism \eqref{transversediffeo} to the general variation \eqref{replicaactioncornervariation} when the metric is on-shell $\overbar{g}_{\mathcal{A}}$ gives
\begin{equation}
\delta_\xi I_{\mathcal{M}_n} = -8\pi\,(1-n)\int_{\mathcal{C}}\hspace{-1mm}d^{D-2}x\sqrt{\sigma}\,(\xi^n\sigma_{ab}\,L^{ab}+\xi^r\sigma_{ab}\,Q^{ab}).
\end{equation}
Requiring the variation to vanish gives the minimal surface condition (\ref{eq:condC}) in $\overbar{g}_{\mathcal{A}}$,
%\begin{equation}
%\sigma^{ab}L_{1ab} =\sigma^{ab}Q_{1ab} = 0.
%\end{equation}
and is equivalent to the condition found by extremizing the area functional in $\overbar{g}_{\mathcal{A}}$ (cf. (\ref{fixedareaentropy}) below). Thus, the area extremization prescription for fixed-area states follows from varying the Einstein action via transverse diffeomorphisms of $\mathcal{C}$. This extremization coincides with the vanishing of the Hamiltonian charge generating $\xi$ \cite{Balasubramanian:2023dpj}.

%\newpage

\vspace{1mm}

\noindent \emph{Entropy functional.} The above two variational principles  determine the on-shell fixed-area metric $\overbar{g}_{\mathcal{A}}$ and the on-shell embedding of $\mathcal{C}$ in $\mathcal{M}$. Using \eqref{replicaactioncorner} on-shell gives
\begin{equation}
I_{\mathcal{M}_n}[\overbar{g}_{\mathcal{A}}] = n\,I_{\mathcal{M}\,\backslash\, \mathcal{C}}[\overbar{g}_{\mathcal{A}}] -4\pi\,(1-n)\,\mathcal{A}.
\label{replicaactioneinstein2}
\end{equation}
where we used $\overbar{g}_{\mathcal{A}} = \overbar{g}_1$ respects replica symmetry. Consequently, the refined R\'enyi entropy \eqref{refinedrenyifixedarea} is
\begin{equation}
\widetilde{S}_n = 4\pi\,\mathcal{A},
\label{fixedareaentropy}
\end{equation}
which is independent of $n$, corresponding to the flat entanglement spectrum of fixed-area states. 
%Further, observe that the non-zero entropy \eqref{fixedareaentropy} comes from the area counterterm included in the wedge action \eqref{fullEinsteinactioncorner}.\JK{This is in slight tension with the abstract.} 
%The fact the coefficient is $4\pi$ (or $1\slash (4G_{\text{N}})$ in the standard normalization) is fixed uniquely by the requirement that the corner term vanishes at $n = 1$ as explained above.

\noindent \textbf{Discussion.} Using Hayward corner terms, we developed a first principles prescription to compute gravitational R\'enyi entropy of Hartle--Hawking and fixed--area states in Einstein gravity.
%for Euclidean solutions to Einstein gravity.\JK{simply: in Einstein gravity?} 
%Thus far we have computed gravitational entropies. 
Via AdS/CFT duality, where the bulk is taken to be asymptotically AdS, our bulk computations directly translate to (refined) R\'enyi entropies of holographic CFTs. When $n\to1$, we recover the Ryu--Takayanagi relation (\ref{eq:RTform}) for Hartle--Hawking states, and the analog for states of fixed--area/flat entanglement spectrum \cite{Fursaev:2006ih}. 
%Previous work used a Hayward term to construct entropy functionals of fixed--area states \cite{Takayanagi:2019tvn}, and R\'enyi entropy in Einstein and Jackiw--Teitelboim gravity \cite{Botta-Cantcheff:2020ywu,Arias:2021ilh}. In contrast, however, rather than varying the corner term, we work directly with the action of the wedge. An appealing feature of this approach, is that our method readily extends to higher-curvature theories of gravity.
Previous work used the Hayward term to construct entropy functionals of fixed--area states \cite{Takayanagi:2019tvn}, and R\'enyi entropy in Einstein and Jackiw--Teitelboim gravity \cite{Botta-Cantcheff:2020ywu,Arias:2021ilh}, but a complete consideration of variational problems for the metric and the corner embedding was lacking. Our formulation based on \eqref{cutHH} and \eqref{cutreplica} allows for a careful treatment of the variational principle. Another appealing feature of our approach is that it readily extends to higher-curvature theories of gravity \cite{Kastikainen:2023omj}.

For example, for Lovelock gravity \cite{Lovelock:1971yv}, which has a known corner term \cite{Cano:2018ckq}, the entropy functional of HH states is given by the Jacobson--Myers (JM) entropy \cite{Jacobson:1993xs}, even when extrinsic curvatures are present. The variational principle of the wedge action is well-posed, giving Lovelock's field equations, which provide a constraint for $\mathcal{C}$ coinciding with extremization of the JM functional. Moreover using the Lovelock corner term, fixed--area states generalize to fixed-JM states, where the Jacobson–-Myers functional of $\mathcal{C}$ is fixed. Thus, the entropy functional of a fixed-JM state is
the Jacobson–-Myers functional with a flat spectrum. The extremization
prescription arises from the variation of the Lovelock action of the wedge and coincides with the extremization of the JM functional. For arbitrary  $F$(Riemann) gravity, corner terms are not known to exist assuming Dirichlet boundary conditions alone (this is also the case for GHY-like terms, cf. \cite{Dyer:2008hb,Deruelle:2009zk,Guarnizo:2010xr,Teimouri:2016ulk,Liu:2017kml}).
%Hence, the variational problem in the presence of corners is not well-posed, and our method cannot provide a generalization of fixed--area states for such theories.
Hence, the variational problem in the presence of corners is not well-posed, and our method suggests fixed--area state analogs do not exist in such theories. 
For HH states, under special periodic boundary conditions, we can recover the Dong--Lewkowycz entropy \cite{Dong:2017xht}, though it has not been proven if this is equal to the Camps--Dong prescription \cite{Dong:2013qoa,Camps:2013zua}. We explain this extension and its limitations in \cite{Kastikainen:2023omj}.

It is worth emphasizing the refined gravitational R\'enyi entropy (\ref{eq:refinedRenyi}) obeys an area-law analogous to the Ryu-Takayanagi relation \cite{Dong:2016fnf}. In particular, for Hartle–Hawking states, the refined entropy is equal to the area of a codimension-2 cosmic brane  minimally coupled to Einstein gravity that backreacts on the ambient geometry by creating a conical deficit \cite{Vilenkin:1981zs}. The brane action is of Nambu-Goto form with a R\'enyi index $n$-dependent tension, $T_{n}=4\pi\,(n-1)/n$ (in units $(16\pi G_{\text{N}})^{-1}=1$), such that in the tensionless limit $n\to1$ the cosmic brane becomes a probe brane, no longer backreacts, and settles at the location of minimal surface, thereby recovering the RT formula. In the proposal of \cite{Dong:2016fnf}, adding a cosmic brane was partially motivated by gravitational duals of modular Hamiltonians \cite{Jafferis:2014lza,Jafferis:2015del}. Our work shows that including a cosmic brane \cite{Dong:2016fnf} is in fact a requirement in order for the variational problem with corners obeying Dirichlet boundary conditions to be well-posed, and amounts to include a Hayward corner term. Indeed, the Hayward term (\ref{fullEinsteinactioncorner}) is of the same form of the cosmic brane action, however, with `tension' $T_{n}=4\pi\,(n-1)$, because we consider solutions with conical excesses, not deficits.

There are multiple possible applications of our work. For example, dynamical black holes are spacetimes without $U(1)$ symmetry and the Camps--Dong formula, at least for linear non-stationary perturbations to stationary horizons, has been argued to be a candidate formula for dynamical black hole entropy \cite{Wall:2015raa}. It would be worth extending our analysis to see if dynamical black hole entropy follows from a corner term. Moreover, it would be worth extending the corner method to the covariant setting \cite{Hubeny:2007xt}, where both timelike and spacelike Hayward terms are needed.

Additionally, our analysis has been at the classical level. It would be interesting to generalize our approach to include bulk quantum corrections, along the lines of \cite{Faulkner:2013ana,Dong:2017xht}. With quantum corrections, our corner method may provide further insight into derivations of the island prescription -- a rule used to compute fine-grained entropy of Hawking radiation \cite{Penington:2019npb,Almheiri:2019psf,Almheiri:2019yqk,Almheiri:2019hni}. In particular, the island rule was derived for Jackiw--Teitelboim gravity using the aforementioned cosmic brane method \cite{Almheiri:2019qdq,Colin-Ellerin:2021jev}, or, alternatively, by computing the microcanonical partition function in \cite{Pedraza:2021ssc,Svesko:2022txo}.

Lastly, a distinct advantage of our approach is that we derived the distributional nature of the Einstein-Hilbert term without producing higher-order regularization terms. Thus, it may be worth revisiting the regularization procedure developed to analyze integrals of curvature invariants on manifolds with a squashed conical excess \cite{Fursaev:2013fta}, and its applications, e.g., quantum corrections to entanglement entropy. We leave these extensions for future work.

%Moreover, the island prescription \cite{Penington:2019npb,Almheiri:2019psf,Almheiri:2019yqk,Almheiri:2019hni} -- used to compute fine-grained entropy of Hawking radiation -- was derived for Jackiw--Teitelboim gravity using the aforementioned cosmic brane method \cite{Almheiri:2019qdq,Colin-Ellerin:2021jev}. Our corner method may provide further insight into these derivations of the island rule. It would also be interesting to compare our corner method to an alternative derivation of the island prescription \cite{Pedraza:2021ssc,Svesko:2022txo} where one instead computes the microcanonical partition function. 

%Lastly, our analysis has been at the classical level. It would be interesting to generalize our approach to include bulk quantum corrections, along the lines of \cite{Faulkner:2013ana,Dong:2017xht}. Moreover, it would be worth extending the corner method to the covariant setting \cite{Hubeny:2007xt}, where both timelike and spacelike Hayward terms are needed. 

\noindent \emph{Acknowledgements.} We are grateful to Pablo Bueno, Luca Ciambelli, Elena C\'{a}ceres, Xi Dong, and Alejandro Vilar L\'{o}pez for useful correspondence. We thank Manus Visser for many fruitful discussions and initial collaboration. JK is supported by the Deutsche Forschungsgemeinschaft (DFG, German Research Foundation) under Germany’s Excellence Strategy through the Würzburg-Dresden
Cluster of Excellence on Complexity and Topology in Quantum Matter - ct.qmat (EXC
2147, project-id 390858490), as well as through the German-Israeli Project Cooperation (DIP) grant ‘Holography and the Swampland’. JK thanks the Osk. Huttunen foundation and the Magnus Ehrnrooth foundation for support during earlier stages of this work. AS is supported by STFC consolidated grant ST/X000753/1 and was partially supported by the Simons Foundation via \emph{It from Qubit Collaboration} and by EPSRC when this work was initiated. AS thanks the Isaac Newton Institute for Mathematical Sciences, Cambridge, for support and hospitality during the program \emph{Black holes: bridges between number theory and holographic quantum information} (supported by EPSRC grant no EP/R014604/1) as this work was being completed.

\bibliographystyle{apsrev4-2}
\bibliography{higher_curvature}

\end{document}